\documentclass[a4paper]{jpconf}
\usepackage{graphicx}

\begin{document}
\title{Robertson-Schr\"odinger formulation of Ozawa's Uncertainty Principle}

\author{Catarina Bastos\footnote[1]{Based on a talk presented by CB at DICE 2014, Castiglioncello, Italy, 15th-19th September 2014}}
\address{Instituto de Plasmas e Fus\~ao Nuclear, Instituto Superior T\'ecnico Avenida Rovisco Pais 1, 1049-001 Lisboa, Portugal}
\ead{catarina.bastos@ist.utl.pt}
\author{Alex E. Bernardini}
\address{Departamento de F\'{\i}sica, Universidade Federal de S\~ao Carlos, PO Box 676, 13565-905, S\~ao Carlos, SP, Brasil.}
\ead{alexeb@ufscar.br}
\author{Orfeu Bertolami}
\address{Departamento de F\'isica e Astronomia, Faculdade de Ci\^encias da Universidade do Porto, Rua do Campo Alegre, 687,4169-007 Porto, Portugal}
\ead{orfeu.bertolami@fc.up.pt}
\author{Nuno Costa Dias and Jo\~ao Nuno Prata\footnote[2]{Also at
Grupo de F\'{\i}sica Matem\'atica, UL, Avenida Prof. Gama Pinto 2, 1649-003, Lisboa, Portugal.}}
\address{Departamento de Matem\'{a}tica, Universidade Lus\'ofona de Humanidades e Tecnologias Avenida Campo Grande, 376, 1749-024 Lisboa, Portugal}
\ead{ncdias@meo.pt, joao.prata@mail.telepac.pt}

\begin{abstract}
A more general measurement disturbance uncertainty principle is presented in a Robertson-Schr\"odinger formulation. It is shown that it is stronger and having nicer properties than Ozawa's uncertainty relations. In particular is invariant under symplectic transformations. One shows also that there are states of the probe (measuring device) that saturate the matrix formulation of measurement disturbance uncertainty principle. 
\end{abstract}

\section{Introduction}

In this work one presents a matrix formulation of Ozawa's measurement disturbance uncertainty principle (OUP), which takes into account position-momentum correlations \cite{Bastos0}. This type of formulation is to the OUP \cite{Ozawa,Ozawa4} as Robertson-Schr\"odinger formulation of uncertainty principle is to the Robertson-like inequality. The latter is usually associated to a set of kinematical restrictions that can be imposed to a quantum system,\begin{equation}
\sigma (A, \psi) \sigma (B, \psi) \ge \frac{| \langle \psi | ~ \left[A, B \right] ~ | \psi\rangle |}{2}~,
\label{eq1}
\end{equation}
where $\sigma (C, \psi)$ denotes the standard mean deviation for an observable $C$: $\sigma^2 (C, \psi) =\langle \psi |(C-\langle \psi |C| \psi\rangle)^2 | \psi\rangle$. Here, $\left[A,B \right] =AB-BA$ is the commutator and $\left\{A,B \right\} = \frac{AB+BA}{2}$ is the anti-commutator. These uncertainty relations can be written in terms of the positivity of a matrix, the covariance matrix of the quantum state ${\bf \Sigma}$, the so-called Robertson-Schrödinger uncertainty principle (RSUP)
\begin{equation}
{\bf \Sigma} + {i \hbar\over2} {\bf J} \ge 0~.
\label{eq3}
\end{equation}
Let $A=X$ be the position of the particle and $B=P$, its momentum, then ${\bf \Sigma}$ is given by 
\begin{equation}
{\bf \Sigma} = \left(
\begin{array}{c c}
\sigma (X, \psi)^2 & \sigma (X,P, \psi)\\
\sigma (P, X, \psi) & \sigma (P, \psi)^2
\end{array}
\right)~,
\label{eq4}
\end{equation}
where $\sigma (X,P, \psi)=\sigma (P,X, \psi) = \langle \psi | ~ \left\{ \Delta X, \Delta P \right\}~ | \psi\rangle $ are the covariance elements for position-momentum correlations, and
\begin{equation}
{\bf J} = \left(
\begin{array}{c c}
0 & 1\\
-1 & 0
\end{array}
\right)
\label{eq5}
\end{equation}
is the standard symplectic matrix.

Why this matrix formulation should be better than the inequality (\ref{eq1})? In fact, the RSUP formulation, Eq. (\ref{eq3}), has several advantages. It implies inequality (\ref{eq1}), thus it is stronger. However, the converse is not true and the RSUP accounts for the correlations between position and momentum. For Gaussian states, the RSUP is a necessary and sufficient condition of quantumness, and it also establishes the separability or entanglement of 2-mode Gaussian states, when a reflection transformation is done \cite{Simon1}. Finally, in contrast with inequalities (\ref{eq1}), the RSUP is invariant under symplectic transformations. An experimental violation of the uncertainty principles (\ref{eq1}) and (\ref{eq3}) it would imply two possibilities. The violation should be due to the failure of the Hilbert space in correctly describing quantum systems, which clearly would have important implications for the theoretical structure of quantum mechanics. Alternatively, it could be a consequence of a deformation of the algebra defined by the pair of observables, position and momenta. In a recent work \cite{Bastos1}, this kind of deformation was studied, even though is no evidence that such an experimental violation has ever been observed. 

When considering measurement-disturbance uncertainty principles, one can relate the accuracy and disturbance in the measurement of a pair of observables. These type of relations are known as error-disturbance trade-off relations and are dynamical uncertainty principles. An example of this relation was related by Heisenberg in his well-known $\gamma$-ray experiment \cite{Heisenberg}, where he obtained a relation between the accuracy of an appropriate position measurement, $\epsilon (X, \psi) $ and the disturbance of the particle+s momentum, $\chi (P, \psi)$,
\begin{equation}
\epsilon (X, \psi) \chi (P, \psi) \ge \frac{|\langle \psi | ~ \left[X, P \right] ~ | \psi\rangle|}{2}~.
\label{eq6}
\end{equation}
Recently, there has been a heated discussion about the status and the interpretation of this Heisenberg's formulation of uncertainty principle \cite{Ozawa,Ozawa4,Busch,Fujikawa,Cyril1}. In fact, experimental violations of this inequality have been found using or weak measurements and polarized entangled photons \cite{Rozema}, or through neutron spin measurements \cite{Hasegawa}, or even through approximate joint measurements of incompatible polarizations observables on single photons \cite{Cyril2}. Theoretically, the different between the two main approaches is based on the definition of the error and disturbance quantities. One one hand, Busch et al. \cite{Busch} considered the measures of error and disturbance as state independent. On the other hand, others considered state dependent quantities \cite{Ozawa,Cyril1, Korzekwa}. The most discussed approach is the OUP, which defines $\epsilon(A, \psi)$ and $\chi(B, \psi)$ as the noise in the measurement of observable $A$ and the disturbance of $B$ due to that measurement, respectively. He considers a composite system of an object and a probe, and initially prepared in a product state $\Psi = \psi \otimes \xi$, where $\psi$ and $\xi$ describe the object and the probe, respectively.
In the Heisenberg picture, a noise operator $N(A)$ associated with observable $A$ and a disturbance operator $D(B)$ have been introduced. They are self-adjoint operators, defined by
\begin{equation}
N(A) = M^{out} - A^{in}\hspace{0.2cm}, \hspace{0.2cm} D(B) = B^{out} - B^{in}~,
\label{eq7}
\end{equation}
where $A^{in}=A \otimes I, B^{in}=B \otimes I$ are observables $A,B$ prior to the measurement interaction, $B^{out} =U^{\dagger} (B \otimes I) U$ is the observable $B$ immediately after the measurement and $M$ is the probe observable. $U$ is the unitary time evolution operator during the measuring interaction. Clearly, $M^{in} = I \otimes M$ and $M^{out} = U^{\dagger} (I \otimes M) U$ \cite{Ozawa4}. Ozawa argued that the noise $\epsilon (A, \psi)$ and disturbance $\eta (B; \psi)$ are defined by:
\begin{equation}
\epsilon (A, \psi)^2 = \langle \Psi | N(A)^2 | \Psi\rangle\hspace{0.2cm} , \hspace{0.2cm} \eta (B, \psi)^2 =  \langle\Psi | D(B)^2 | \Psi\rangle~.
\label{eq8}
\end{equation}
Then, as $M$ and $B$ are observables in different systems, they commute $\left[M^{out}, B^{out} \right]=0$. Thus, using these last definitions, the triangle and Cauchy-Schwartz inequalities one has finally the OUP,
\begin{equation}
\epsilon (A, \psi) \eta (B, \psi) +\epsilon (A, \psi) \sigma (B, \psi) +  \sigma (A, \psi) \eta (B, \psi) \ge {{| \langle\psi| ~\left[A, B \right] ~| \psi\rangle}\over2}~.
\label{eq11}
\end{equation}
Clearly, this relation does not take into account noise-disturbance correlations. Thus, in what follows one shows how to obtain a more general and stronger formulation of this OUP and where one has information about these correlations.

\section{Matrix Formulation of Ozawa Uncertainty Principle}

One considers a multidimensional system, where $A_i^{in}$ and $B_j^{in}$, $i,j=1, \cdots, n$ such that
\begin{equation}
\left[A_i^{in},A_j^{in} \right] = \left[B_i^{in},B_j^{in} \right] =0\hspace{0.2cm}, \hspace{0.2cm} \left[A_i^{in}, B_j^{in} \right] = i C_{ij}~,
\label{eq12}
\end{equation}
for $i,j=1, \cdots, n$, and where $\left\{C_{ij} \right\}$ are some self-adjoint operators. If $A$ and $B$ are the particle's position and momentum, one simply has $C_{ij} = \hbar \delta_{ij} $. These self adjoint operators can be collectively written as $Z^{in} = \left(A_1^{in}, \cdots, A_n^{in},B_1^{in}, \cdots, B_n^{in} \right)$ satisfying the commutation relations
\begin{equation}
\left[Z_{\alpha}^{in},Z_{\beta}^{in} \right] = i G_{\alpha \beta}\hspace{0.2cm}, \hspace{0.2cm} \alpha, \beta =1, \cdots, 2n~.
\label{eq14}
\end{equation}
${\bf G} = \left\{G_{\alpha \beta} \right\}$ is the self-adjoint operator-valued skew-symmetric matrix
\begin{equation}
{\bf G} = \left(
\begin{array}{c c}
{\bf 0} & {\bf C}\\
- {\bf C} & {\bf 0}
\end{array}
\right)~,
\label{eq15}
\end{equation}
with ${\bf C} = \left\{C_{ij} \right\}$. Again, if $A$ and $B$ are the position and momentum operators, then one simply has $2n \times 2n$ standard symplectic matrix $\hbar {\bf J}$. The same strategy can de used to write collectively the noise and disturbance operators,
\begin{eqnarray}
N  = N (A) = \left(N_1, \cdots, N_n \right)\hspace{0.2cm},\hspace{0.2cm}D  = D (B) = \left(D_1, \cdots, D_n \right)~,
\label{eq17}
\end{eqnarray}
into $K = \left(N_1, \cdots, N_n, D_1, \cdots, D_n \right)$. Finally, let 
\begin{eqnarray}
M^{out} = \left(M^{out}_1, \cdots, M^{out}_n \right)\hspace{0.2cm},\hspace{0.2cm}
B^{out} = \left( B_1^{out}, \cdots, B_n^{out} \right)~,
\label{eq20}
\end{eqnarray}
denote the output of the (commuting) probe observables and the output of $B$, respectively. They can be written collectively into $Z^{out} = \left(M^{out}_1, \cdots, M^{out}_n,B_1^{out}, \cdots, B_n^{out} \right)$.
Thus, as before one has $\left[Z_{\alpha}^{out}, Z_{\beta}^{out} \right]=0\hspace{0.2cm}, \hspace{0.2cm} \alpha , \beta =1, \cdots, 2n$ and $Z^{out}= Z^{in} + K$.

Considering an arbitrary set of complex numbers denoted by $\left\{\lambda_{\alpha} \right\}_{1 \le \alpha \le 2n}$, one has \cite{Bastos0}
\begin{eqnarray}
0 &=& \sum_{\alpha, \beta =1}^{2n} \overline{\lambda}_{\alpha} \lambda_{\beta} \langle \left[\Delta Z_{\alpha}^{out},  \Delta Z_{\beta}^{out} \right]\rangle \nonumber\\
&=&\sum_{\alpha, \beta =1}^{2n} \overline{\lambda}_{\alpha} \lambda_{\beta} \left(i \langle G_{\alpha \beta}\rangle + \langle \left[ Z_{\alpha}^{in},K_{\beta} \right]+  \left[ K_{\alpha} ,Z_{\beta}^{in} \right]\rangle +  \langle \left[K_{\alpha} ,K_{\beta} \right]\rangle\right)~.
\label{eq24}
\end{eqnarray}
Writing $\mathcal{K} = \sum_{\alpha}\lambda_{\alpha} K_{\alpha}$, and assuming that $\mathcal{K} \mathcal{K}^{\dagger}$ is a positive operator, one has
\begin{equation}
\sum_{\alpha, \beta =1}^{2n} \overline{\lambda}_{\alpha} \lambda_{\beta}  \langle \left[ K_{\alpha} , K_{\beta} \right]\rangle = \langle \mathcal{K}^{\dagger} \mathcal{K} \rangle- \langle\mathcal{K} \mathcal{K}^{\dagger} \rangle ~\le ~  \langle\mathcal{K}^{\dagger} \mathcal{K}\rangle~.
\label{eq25}
\end{equation}
Thus,
\begin{equation}
\sum_{\alpha, \beta =1}^{2n} \overline{\lambda}_{\alpha} \lambda_{\beta}  \langle \left[ K_{\alpha} , K_{\beta} \right]\rangle ~ \le ~ \sum_{\alpha, \beta =1}^{2n} \overline{\lambda}_{\alpha} \lambda_{\beta} \langle  K_{\alpha}  K_{\beta}\rangle=\sum_{\alpha, \beta =1}^{2n} \overline{\lambda}_{\alpha} \lambda_{\beta}\left( \langle \left\{ K_{\alpha} , K_{\beta} \right\}\rangle + {1\over 2} \langle \left[ K_{\alpha} , K_{\beta} \right]\rangle \right)~.
\label{eq26}
\end{equation}
And so, 
\begin{equation}
\sum_{\alpha, \beta =1}^{2n} \overline{\lambda}_{\alpha} \lambda_{\beta}  \langle \left[K_{\alpha} ,K_{\beta} \right]\rangle ~ \le ~ 2 \sum_{\alpha, \beta =1}^{2n} \overline{\lambda}_{\alpha} \lambda_{\beta}\langle \left\{ K_{\alpha} , K_{\beta} \right\}\rangle~.
\label{eq27}
\end{equation}
Upon substitution of Eq. (\ref{eq27}) into Eq. (\ref{eq24}) one finally obtains
\begin{equation}
0 \le \sum_{\alpha, \beta =1}^{2n} \overline{\lambda}_{\alpha} \lambda_{\beta} \left(i \langle G_{\alpha \beta}\rangle + \langle \left[ Z_{\alpha}^{in},K_{\beta} \right] +  \left[K_{\alpha} ,Z_{\beta}^{in} \right]\rangle +  2 \langle \left\{K_{\alpha} , K_{\beta} \right\}\rangle\right)~.
\label{eq28}
\end{equation}
Then, defining the $2n \times 2n$ real symmetric positive-definite matrix
\begin{equation}
{\bf K}_{\alpha \beta} =   \langle \left\{K_{\alpha} ,K_{\beta} \right\}\rangle~,
\label{eq29}
\end{equation} 
the $2n \times 2n$ real skew-symmetric matrices ${\bf \mathcal{G}} = \langle{\bf G}\rangle$ and
\begin{equation}
{\bf \Gamma}_{\alpha \beta} ={1\over i}  \langle \left[ Z_{\alpha}^{in},K_{\beta} \right] +  \left[K_{\alpha} ,Z_{\beta}^{in} \right]\rangle~,
\label{eq30}
\end{equation} it is possible to rewrite (\ref{eq28}) in the matrix form:
\begin{equation}
{\bf K} + {i\over2} \left( {\bf \Gamma} + {\bf \mathcal{G}} \right) \ge 0~,
\label{eq31}
\end{equation}
which is a matrix version of OUP. If a measuring interaction is of independent intervention, then ${\bf \Gamma} =0$, and one recovers the matrix generalization of Heisenberg's noise disturbance relation Eq. (\ref{eq6}),
\begin{equation}
{\bf K} + {i\over 2} {\bf \mathcal{G}}  \ge 0~.
\label{eq33}
\end{equation}

Now one shows that the uncertainty principle, Eq. (\ref{eq31}), is in fact stronger than the OUP. In the case of $n=1$, one finds
\begin{equation}
{\bf K} = \left(
\begin{array}{c c}
\langle N^2\rangle & \langle\left\{ N, D \right\}\rangle \\
\\
\langle\left\{D, N \right\}\rangle & \langle D^2\rangle
\end{array}
\right)\hspace{0.2cm},\hspace{0.2cm}
{\bf \Gamma} = {1\over i} \left(
\begin{array}{c c}
0 & \langle\left[ A^{in},  D \right] +\left[N, B^{in} \right] \rangle \\
\\
\langle\left[  D , A^{in}\right] +\left[ B^{in}, N   \right] \rangle & 0
\end{array}
\right)~.
\label{eq35}
\end{equation}
Clearly, if inequality (\ref{eq31}) holds, then the matrix ${\bf K} + {i\over2} \left( {\bf \Gamma} + {\bf \mathcal{G}} \right)$ must have non-negative determinant, and then \cite{Bastos0}
\begin{eqnarray}
\langle N^2\rangle \langle D^2\rangle   &\ge& \langle\left\{ N, D \right\}\rangle^2+{1\over4} \left| \langle\left[ A^{in},  D \right] +\left[N, B^{in}   \right] \rangle + \langle \left[A^{in}, B^{in} \right]\rangle \right|^2\nonumber\\
 &\ge& {1\over4} \left| \langle\left[ A^{in},  D \right] +\left[N, B^{in}   \right]  \rangle + \langle \left[A^{in}, B^{in} \right]\rangle \right|^2~.
 \label{eq36}
\end{eqnarray}
Taking the square root of this last inequality,
\begin{equation}
\langle N^2\rangle^{1/2} \langle D^2\rangle^{1/2} \ge  {1\over2} \left| \langle\left[ A^{in},  D \right] +\left[N, B^{in}   \right] \rangle - \langle \left[B^{in}, A^{in} \right]\rangle \right|~,
\label{eq37}
\end{equation}
considering the definitions $\epsilon (A) = \langle N^2\rangle^{1/2}$ and $\eta (B) = \langle D^2\rangle^{1/2}$, and using the inequality $|a-b| \ge \left|~ |a|- |b|  ~ \right|$, one obtains
\begin{equation}
\epsilon (A) \eta (B) \ge  {1\over2} \left| ~| \langle\left[ A^{in},  D \right] +\left[N, B^{in}   \right]  \rangle| - |  \langle \left[A^{in}, B^{in} \right]\rangle  | ~\right|~.
\label{eq38}
\end{equation}
And in particular
\begin{equation}
\epsilon (A) \eta (B) \ge {1\over2} |  \langle \left[A^{in}, B^{in} \right]\rangle  | - {1\over2} \left|  \langle\left[ A^{in},  D \right] +\left[N, B^{in}   \right]  \rangle\right|~,
\label{eq39}
\end{equation}
which is OUP, Eq. (\ref{eq11}).

\section{Properties of the Matrix Formulation of the measurement disturbance uncertainty principle}

In order to demonstrate that the matrix formulation of OUP, inequality (\ref{eq31}),  is in fact stronger than OUP, one uses the example of a backaction evading quadrature amplifier (BAE) \cite{Bastos0,Ozawa4}. The system is described by a set of quadrature operators $X_a,Y_a$ and the probe by the operators $X_b,Y_b$ with
\begin{equation}
\left[X_a,Y_a \right]= \left[X_b,Y_b \right] = {i\over2}~.
\label{eq40}
\end{equation}
Then, the measuring interaction is given by:
\begin{equation}
\left\{
\begin{array}{l}
X_a^{out}= X_a^{in}\\
X_b^{out} = X_b^{in} +  G X_a^{in}\\
Y_a^{out} = Y_a^{in} - G Y_b^{in}\\
Y_b^{out} = Y_b^{in}~,
\end{array}
\right.
\label{eq41}
\end{equation}
where $G$ is the gain. The probe observable is then set to $M= {1\over G} X_b$, and thus
\begin{equation}
M^{out} = X_a^{in} + {1\over G} X_b^{in}~.
\label{eq42}
\end{equation}
Moreover
\begin{equation}
N (X_a)= {1\over G} X_b^{in}\hspace{0.1cm},\hspace{0.1cm}D (X_a) =0\hspace{0.1cm},\hspace{0.1cm}D (Y_a) = - G Y_b^{in}
\label{eq43}
\end{equation}
or in the notation introduced above
\begin{equation}
K = \left(  {1\over G} X_b^{in}, -G Y_b^{in} \right)~.
\label{eq45}
\end{equation}
Clearly, ${\bf \Gamma } =0$ and the measuring interaction is of independent intervention for the pair $(X_a,Y_a)$. Also from Eq. (\ref{eq40}), one finds ${\bf \mathcal{G}} = {1\over 2} {\bf J}$.

One considers the state of the probe $\xi$ to have vanishing expectation values $\langle\xi|X_b^{in}|\xi\rangle=\langle\xi|Y_b^{in}|\xi\rangle=0$, and a covariance matrix
\begin{equation}
{\bf \Sigma^b} = \left(
\begin{array}{c c}
{\bf \Sigma^b}_{11} & {\bf \Sigma^b}_{12} \\
& \\
{\bf \Sigma^b}_{12} & {\bf \Sigma^b}_{22}
\end{array}
\right) =
\left(
\begin{array}{c c}
1/4 & 1/2 \\
& \\
1/2 & 1/4
\end{array}
\right)~.
\label{eq47}
\end{equation}
One obtains the elements of $\bf K$ from Eq. (\ref{eq45}) \cite{Bastos0}. For instance, 
\begin{equation}
\langle N^2\rangle = {1\over G^2} \langle ( X_b^{in})^2\rangle = {1\over G^2} {\bf \Sigma^b}_{11} =\frac{1}{4 G^2}~.
\label{eq48}
\end{equation}
The other elements are obtained in an analogous way. Thus,
\begin{equation}
{\bf K} = \left(
\begin{array}{c c}
{1\over 4 G^2}  & -{1\over2}\\
& \\
-{1\over2} & {G^2\over4}
\end{array}
\right)~.
\label{eq51}
\end{equation}
We then have from ${\bf \mathcal{G}} = {1\over 2} {\bf J}$ and Eq. (\ref{eq51}):
\begin{equation}
{\bf K} + {i\over2} \left({\bf \Gamma} + {\bf \mathcal{G}} \right)={\bf K} + {i\over4} {\bf J} = \left(
\begin{array}{c c}
{1\over 4 G^2}  & - {1\over2} + {i\over4}\\
& \\
-{1\over2} -{i\over4} &  {G^2\over4}
\end{array}
\right)~.
\label{eq52}
\end{equation}
Since $\det ({\bf K} + {i\over4} {\bf J}) = - {1\over4}$, then it is not a positive matrix and the uncertainty principle (\ref{eq31}) is violated. However, since $\epsilon (A) = \langle N^2 \rangle^{1/2}= {1\over2G}$ and $\eta (B) = \langle D^2 \rangle^{1/2}= {G\over2}$, one concludes that $\epsilon (A)\eta (B) ={1\over4}$, which exactly saturates the OUP. This shows that in fact the inequality (\ref{eq31}) is indeed stronger that the OUP.

Next, one shows that the uncertainty principle, Eq. (\ref{eq31}), can be saturated if there is a set of complex numbers $\left\{\lambda_{\alpha}\right\}_{1 \le \alpha \le 2n}$ and a state $\Psi$ such that 
\begin{equation}
\sum_{1 \le \alpha \le 2n} \lambda_{\alpha} K_{\alpha} \Psi =0~.
\label{eq33.1}
\end{equation}
This may or may not be possible, depending on the commutation relations of the operators $\left\{K_{\alpha}\right\}_{1 \le \alpha \le 2n}$. In the case of BAE described above this is not possible. Indeed, to have a saturating state $\Psi$, Eq. (\ref{eq33.1}), would have to satisfy:
\begin{equation}
\left( {{\lambda_1}\over G} X_b^{in} +{i\over2} G \lambda_2 {\partial\over{\partial X_b^{in}}} \right) \Psi =0~.
\label{eq33.4}
\end{equation}
for some constants $(\lambda_1, \lambda_2 ) \in {C}^2 \backslash \left\{(0,0) \right\}$. In the representation of quadrature operators and using their commutation relations, one concludes that there are no square-integrable solutions, and thus there is no saturating state \cite{Bastos0}.

Alternatively, if one considers a noiseless quadrature transducer \cite{Ozawa4} with the following input-output relations for the quadrature operators (\ref{eq40})
\begin{equation}
\left\{
\begin{array}{l}
X_a^{out}= X_a^{in}- X_b^{in}~,\\
X_b^{out}= X_a^{in}~,\\
Y_a^{out}=- Y_b^{in}~,\\
Y_b^{out}= Y_b^{in}+ Y_a^{in}~.
\end{array}
\right.
\label{eq33.5}
\end{equation}
and a probe observable $M=X_b$, then
\begin{equation}
K=(0,-Y_a^{in} - Y_b^{in})~.
\label{33.7}
\end{equation}
And thus, a saturating state satisfies
\begin{equation}
- \lambda_2  (Y_a^{in} + Y_b^{in}) \Psi =0~.
\label{eq33.8}
\end{equation}
This can be achieved by any state $\Psi$ for $\lambda_2 =0$.

Finally, one should notice that the matrix formulation of the noise-disturbance uncertainty principle, Eq. (\ref{eq31}) remains unchanged under symplectic transformations. 
To show that one considers that $\left[ K_{\alpha} , K_{\beta} \right] = i \gamma J_{\alpha \beta}$, where $\gamma \ne 0$ is some real constant and ${\bf J}$ is the standard symplectic matrix. Suppose like in the BAE casa that (${\bf \Gamma} =0$). Then Eq. (\ref{eq31}) becomes ${\bf K} + {{i \gamma}\over2} {\bf J} \ge 0$. Notice that this looks formally like the RSUP Eq. (\ref{eq3}). If the system undergoes a linear symplectic transformation as $K_{\alpha} \mapsto K_{\alpha}^{\prime} = \sum_{1 \le \beta \le 2n} S_{\alpha \beta} K_{\beta}$, where ${\bf S} \in Sp(2n; {R})$, then the noise-disturbance covariance matrix transforms by similarity ${\bf K} \mapsto {\bf K}^{\prime} = {\bf S}{\bf K} {\bf S}^T$. Since ${\bf S}^{-1} {\bf J} ({\bf S}^{-1})^T = {\bf J}$, one concludes that the matrix uncertainty principle remains unchanged: ${\bf K}^{\prime} + {{i \gamma}\over2} {\bf J} \ge 0$.

In contrast, the OUP, when considering for instance the previous example of the BAE, becomes simply the Heisenberg inequality
\begin{equation}
\epsilon \eta \ge{1\over4}~.
 \label{eq53}
 \end{equation}
Rewriting this now as a noise-disturbance covariance matrix ${\bf K} = diag (\epsilon^2,\eta^2)$ and considering that the probe (and possibly the object) is subjected to a symplectic transformation such that:
\begin{equation}
K \mapsto K^{\prime} = {\sqrt{2}\over2} \left(
\begin{array}{c c}
1 & 1\\
-1 & 1
\end{array}
\right) K~.
\label{eq54}
\end{equation}
One clearly sees that the noise-disturbance vector $K$ is rotated through an angle of ${\pi\over4}$. Such a transformation can be implemented by a certain unitary transformation $U(S)$ generated by an appropriate hermitian operator, quadratic in the variables $X_b, Y_b$ of the probe. Thus, the OUP is modified to
\begin{equation}
(\epsilon^{\prime} )^2 + (\eta^{\prime} )^2  \ge \sqrt{ 1 + 4 \langle \left\{ N^{\prime} , D^{\prime} \right\} \rangle^2},
\label{eq55}
\end{equation}
which is manifestly different from Eq. (\ref{eq53}).

\section{Conclusions}
One has presented here an universal matrix formulation of the measurement disturbance uncertainty principle. It is more stringent than the Ozawa's measurement disturbance relation as it in fact implies the OUP. Then one has showed that there are states that saturate the OUP but violates the matrix formulation of it. Furthermore, it is possible to obtain states that saturate also the matrix formulation of OUP. 
Finally, it is worth to mention that it could be interesting to investigate the experimental validity of this more general formulation of measurement disturbance uncertainty relation. In fact, the OUP proved to be valid through experimental work performed using weak measurements by Rozema et al. \cite{Rozema} and considering 3-state mode systems by Sulyok et al. \cite{Hasegawa} and by Ringbauer et al. \cite{Cyril2}.

\ack{The work of CB is supported by the FCT (Portugal) grant SFRH/BPD/62861/2009.}

\smallskip

\section*{References}
\begin{thebibliography}{9}

\bibitem{Bastos0} C. Bastos, A.E. Bernardini, O. Bertolami, N.C. Dias and J.N. Prata, Phys. Rev. A 89 (2014) 042112.

\bibitem{Ozawa} M. Ozawa, Phys. Rev. Lett. 60 (1988) 385; Phys. Lett. A 299 (2002) 1; {\it Squeezed and Nonclassical Light}, edited by P. Tombesi and E.R. Pike (Plenum, New York, 1989) pp. 263-268.

\bibitem{Ozawa4} M. Ozawa, Phys. Rev. A 67 (2003) 042105.

\bibitem{Simon1} R. Simon, Phys. Rev. Lett. 84 (2000) 2726.

\bibitem{Bastos1} C. Bastos, O. Bertolami, N.C. Dias, J.N. Prata, Phys. Rev. D 86 (2012) 105030.

\bibitem{Heisenberg} W. Heisenberg, Z. Phys. 43 (1927) 172.

\bibitem{Busch} P. Busch, P. Lahti, R.F. Werner, Phys. Rev. Lett. 111 (2013) 160405; "Noise Operators and Measures of RMS Error and Disturbance in Quantum Mechanics", arxiv: quant-ph/1312.4393.

\bibitem{Fujikawa} K. Fujikawa, Phys. Rev. A 85 (2012) 062117.

\bibitem{Cyril1} C. Branciard, Proc. Nat. Aca. Sci. (USA) 110 (2013) 6742.

\bibitem{Rozema} L. A. Rozema, A. Darabi, D. H. Mahler, A. Hayat, Y. Soudagar, and A. M. Steinberg, Phys. Rev. Lett. 109, 100404 (2012).

\bibitem{Hasegawa} G. Sulyok, S. Sponar, J. Erhart, G. Badurek, M. Ozawa and Y. Hasegawa, Phys. Rev. A 88, 022110 (2013).

\bibitem{Cyril2} M. Ringbauer, D. Biggerstaff, M. Broome, A. Fedrizzi, C. Branciard and A. White, Phys. Rev. Lett. 112 (2014) 020401.

\bibitem{Korzekwa} K. Korzekwa, D. Jennings, T. Rudolph, Phys. Rev. A 89 (2014) 052108.



\end{thebibliography}
\end{document}